\documentclass[prd,tightenlines,nofootinbib,superscriptaddress,showkeys]{revtex4}
\usepackage[utf8x]{inputenc}
\usepackage{amsmath} 
\usepackage{amsfonts,hyperref}
\usepackage{slashed}
\usepackage[dvips]{graphicx}
\usepackage{color}

\allowdisplaybreaks

\newcommand{\und}{{\text{ and }}}

\newcommand{\sgn}{{\text{sgn}}}

\newcommand{\E}{{\text{\textbf{E}}}}
\newcommand{\K}{{\text{\textbf{K}}}}
\newcommand{\cl}{{\text{cl}}}
\renewcommand{\O}{{\mathcal{O}}}
\renewcommand{\Im}{{\text{Im}}}

\newcommand{\e}{{\text{e}}}
\renewcommand{\i}{{\text{i}}}
\renewcommand{\e}{{\text{e}}}

\newcommand{\scE}{{\mathcal{E}}}
\newcommand{\Ds}{{D_s}}
\newcommand{\Eq}{{Eq.~}}
\newcommand{\Eqs}{{Eqs.~}}

\begin{document}

\author{{\bf Eckhard Strobel}\email{eckhard.strobel@gravity.fau.de}}
\affiliation{ICRANet, Piazzale della Repubblica 10, 65122 Pescara, Italy}
\affiliation{Dipartimento di Fisica, Universit\`a di Roma "La Sapienza", Piazzale Aldo Moro 5, 00185 Rome, Italy}
\affiliation{Universit\'e de Nice Sophia Antipolis, 28 Avenue de Valrose, 06103 Nice Cedex 2, France}
\author{{\bf She-Sheng Xue}\email{xue@icra.it}}
\affiliation{ICRANet, Piazzale della Repubblica 10, 65122 Pescara, Italy}
\affiliation{Dipartimento di Fisica, Universit\`a di Roma "La Sapienza", Piazzale Aldo Moro 5, 00185 Rome, Italy}

\date{\today}

\title{Semiclassical pair production rate for time-dependent electrical fields with more than one component: -WKB-approach and  world-line instantons}

%\pacs{12.20.Ds, 11.15.Kc, 11.15.Tk} % QED: -specific calculations, Gauge Field Theories: 11.15.Kc Classical and semiclassical techniques, other nonperturbative techniques

\begin{abstract}
 We present an analytic calculation of the semiclassical electron-positron pair creation rate by time-dependent electrical fields. We use two methods, first the imaginary time method in the WKB-approximation and second the world-line instanton approach. The analytic tools for both methods are generalized to time-dependent electric fields with more than one component.\\ 
  For the WKB method an expansion of the momentum spectrum of produced pairs around the canonical momentum \(\vec{P}=0\) is presented which simplifies the computation of the pair creation rate. We argue that the world-line instanton method of \cite{Dunne2006} implicitly performs this expansion of the momentum spectrum around  \(\vec{P}=0\). Accordingly the generalization to more than one component is shown to agree with the WKB result obtained via this expansion.\\
 However the expansion is only a good approximation for the cases where the momentum spectrum is peaked around \(\vec{P}=0\). Thus the expanded WKB result and the world-line instanton method of \cite{Dunne2006} as well as the generalized method presented here are only applicable in these cases.
 \\
We study the two component case of a rotating electric field and find a new analytic closed form for the momentum spectrum using the generalized WKB method. The momentum spectrum for this field is not peaked around \(\vec{P}=0\).
\end{abstract}
 
 \maketitle

\section*{Introduction}
Since Sauter in 1931 \cite{Sauter1931} and Heisenberg and Euler \cite{Heisenberg1936} four years later gave a first description of the vacuum properties of QED, there have been a lot of investigations of the pair creation rate in strong electric fields. In particular, Schwinger \cite{Schwinger1951,Schwinger1954A,Schwinger1954B} reformulated their result in an elegant way using quantum-field theoretic methods (see also \cite{Nikishov1969,Batalin1970}).\\
The formulation was extended to space-time-dependent fields using different methods, e.g.~the imaginary time method \cite{Brezin1970,Popov1971,Popov1972,Popov1973,Marinov1977,Popov2001} and a tunneling picture \cite{Kleinert2008,Kleinert2013}, both using WKB-approximations or the world-line instanton method \cite{Dunne2005B,Dunne2006}.\\
By comparing numerical with analytic results it was found that for more complicated field configurations, i.e.~for those which have more than one distinct pair of semiclassical turning points, interference effects arise. This was already discussed as a resonance effect for oscillating fields in \cite{Popov1973B}. Interference effects were recently studied in \cite{Dumlu2010,Dumlu2011} for the WKB-method and in \cite{Dumlu2011WLI} for the world-line instanton approach. In this paper we consider only fields with one dominant pair of turning points where interference effects are negligible. This enables us to use scalar quantum-electro dynamics, since it is known to give the same results as spinor quantum electrodynamics at the leading non-perturbative order if there are no interference effects \cite{Dumlu2011}. \\
All the analytic methods mentioned above give the same results for electric fields with only one component depending either on space or time. A more general case, namely electric fields with two or three components depending on space was discussed in \cite{Dunne2006B} in the world-line instanton approach. \\ 
A special case namely a (two component) rotating electrical field was discussed in \cite{Popov1973}. Recently pair production in rotating fields has been studied numerically in \cite{Blinne2013} using the Wigner formalism. These results  can be used to calculate the pair creation rate of a plane wave in a plasma as shown in \cite{Bulanov2003}. \\
So far, electron-positron pair production has not been directly observed in experiments due to the necessity of high field strengths which are out of the range reached by nowadays laser systems. However recent theoretical investigations have shown   that less strong fields are needed if one uses carefully-shaped multi-component laser pulses  \cite{Schutzhold2008,Dunne2009,Bell2008,DiPiazza2009,Monin2010,Monin2010B,Heinzl2010,Bulanov2010}. \\
For this reason, we generalize the above mentioned analytic methods to compute the pair creation rate for a general time-dependent periodic electrical field which is characterized by the potential
\begin{align}
 A_\mu(t)=[0,A_1(t),A_2(t),A_3(t)]=\frac{1}{e\,c}[0,V_1(t),V_2(t),V_3(t)] .\label{eq:potential}
\end{align}
To do so we use the WKB-approximation as well as the world-line instanton method of \cite{Dunne2006}.\\ 
As is well known, the WKB approach the pair creation rate per volume \(V\) takes the general form (see, e.g.,~\cite{Kleinert2008,Kleinert2013}) 
\begin{align}
 \frac{\Gamma_\text{WKB}}{V}&\sim \int \frac{d^3P}{(2\pi\hbar)^3}\exp\left(-\pi\frac{E_c}{E_0}G(\vec{P})\right). \label{eq:2}
\end{align}
 where the integral over \(\vec{P}\) is over the momentum modes of the produced pairs. We introduce the critical electrical field
\begin{align}
 E_c=\frac{m^2c^3}{e \hbar} .
\end{align}
In \Eq(\ref{eq:2}) \(E_0\) is a characteristic electric field strength and \(G(\vec{P})\) is a function depending on the explicit form of the electric field, which is straightforwardly generalized to more than one component.\\
We find that if the momentum spectrum \(\exp(-\pi E_c/E_0G(\vec{P}))\) is peaked around zero canonical momentum \(\vec{P}=0\) it can be approximated by expanding around this point and it is possible to simplify the result via Gaussian integration.\\
In the world-line instanton framework of \cite{Dunne2006} the momentum, arising as an integration constant, was implicitly taken to vanish with a Gaussian momentum integration producing the prefactors, as discussed in \cite{Dumlu2011WLI}. We argue that this is a \textit{de facto} expansion around \(\vec{P}=0\). We generalize this method to  the case of electric fields with more than one component and show that the result agrees with the WKB result expanded around \(\vec{P}=0\).\\
 Looking at examples of electric fields with two components we find that the momentum spectrum is not necessarily peaked around zero momentum. In these cases the expanded WKB result as well as the equivalent  world-line instanton method based on the one of \cite{Dunne2006} cannot be applied, because they do not represent a good approximation.
In general it becomes obvious that to compute the pair creation rate one needs to have knowledge about the momentum spectrum.\\ 
This paper is arranged as follows. In Section \ref{sec:WKB} we compute the pair creation rate in the WKB approximation for fields with one to three components. We repeat the same for the world-line instanton approach in Section \ref{sec:WLI}. In Section \ref{sec:compare} we compare the two methods. We study some examples of interest in Section \ref{sec:app}. Section \ref{sec:conclusions} contains our conclusions and remarks. In order to make the text and ideas more transparent, we relegate some of the technical calculations to \ref{app:series}.
In \ref{app:Morse} we study the value of the Morse index which is important for the calculations in the world-line instanton approach.

%%%%%%%%%%%%%%%%%%%%%%%%%%%%%%%%%%%%%%%%%%%%%%%%%%%%%%%%%%%%%%%%%%%%%%%%%%%%%%%%%%%%%%%%%%%%%%%%%%
\section{Pair production rate for electric fields depending on time in the WKB approximation}
%%%%%%%%%%%%%%%%%%%%%%%%%%%%%%%%%%%%%%%%%%%%%%%%%%%%%%%%%%%%%%%%%%%%%%%%%%%%%%%%%%%%%%%%%%%%%%%%%%
\label{sec:WKB}

Here we briefly review the computation of the pair creation rate for time-dependent fields in the WKB-approximation \cite{Brezin1970,Popov1971,Popov1972,Popov1973,Marinov1977,Popov2001,Dumlu2011}.  
In this case the Klein-Gordon equation reduces to an effective Schrödinger equation. The pair creation rate can thus be connected to the reflection coefficient of a scattering problem.\\
In Section \ref{sec:transmission}  we recall the calculation of the WKB momentum spectrum. It depends on integrals between conjugated pairs of complex turning points in analogy to \cite{Dumlu2010,Dumlu2011}. If there is more than one pair of turning points interference effects can occur. These are governed by integrals between these different pairs. For the scope of this paper we will however concentrate on the case of one dominant pair of turning points for which interference is negligible.
This also enables one to use scalar quantum electro dynamics.  Since as shown in \cite{Dumlu2011} for the case of no interference effects the results obtained in this way are equivalent to the ones  of spinor quantum electrodynamics at the leading non perturbative order.
\\
In Section \ref{sec:K} we show how to simplify the integration between the turning points for analytic purposes.
This involves a generalization of a well known variable substitution of the one component case.
The pair production rate can be calculated from the transmission probability via a integration over the momentum spectrum as discussed in Section \ref{sec:paircreation}. 
For the comparison of the WKB results with the world-line instanton method we expand the momentum spectrum around \(\vec{P}=0\). By doing so it is possible to perform a Gaussian integration in the momentum space. 

%%%%%%%%%%%%%%%%%%%%%%%%%%%%%%%%%%%%%%%%%%%%%%%%%%%%%%%%%%%%%%%%%%%%%%%%%%%%%%%%%%%%%%%%%%%%%%%%%%%%
\subsection{Momentum spectrum in the WKB approximation}
%%%%%%%%%%%%%%%%%%%%%%%%%%%%%%%%%%%%%%%%%%%%%%%%%%%%%%%%%%%%%%%%%%%%%%%%%%%%%%%%%%%%%%%%%%%%%%%%%%%%
\label{sec:transmission}

In this Section we shortly recall the computation of the momentum spectrum within in the WKB method \cite{Brezin1970,Popov1971,Popov1972,Popov1973,Marinov1977,Popov2001,Dumlu2011}.
We start from the Klein-Gordon equation
\begin{align}
 \left([\i \hbar \partial_\mu+{e}A_\mu(t)]^2-m^2 c^2\right)\phi(x,t)=0. \label{eq:KG}
\end{align}
where the electromagnetical potential takes the form (\ref{eq:potential}).
Now the scalar field operator can be decomposed as
\begin{align}
 \hat{\phi}(x,t)=\int\frac{d^3 P}{(2\pi\hbar)^3}\e^{\frac{\i}{\hbar}\vec{P} \cdot\vec{x}} \left(\phi_{\vec{P}}(t)\hat{a}_{\vec{P}}+\phi_{\vec{P}}^*(t)\hat{b}^\dagger_{-\vec{P}}\right), \label{eq:ansatz2}
\end{align}
where \(\hat{a}_{\vec{P}}\) and \(\hat{b}^\dagger_{-\vec{P}}\) are bosonic creation and annihilation operators.
The Klein-Gordon equation (\ref{eq:KG}) for the modes becomes
\begin{align}
 \left(-\hbar^2\partial_{t}^2-\left(\scE(t)\right)^2\right)\phi_{\vec{P}}(t)=0, \label{eq:KGmode}
\end{align}
where we define
\begin{align}
 \left(\scE(t)\right)^2=\left[cP_j-V_j(t)\right]^2+m^2c^4. \label{eq:scE}
\end{align}
One can now perform a Bogoliubov transformation to time-dependent creation and annihilation operators  
\begin{align}
 \hat{c}_{\vec{P}}(t)=\alpha_{\vec{P}}(t)\hat{a}_{\vec{P}}+\beta^*_{\vec{P}}(t)\hat{b}^\dagger_{-\vec{P}}\,,&&\hat{d}^\dagger_{-\vec{P}}(t)=\beta_{\vec{P}}(t)\hat{a}_{\vec{P}}+\alpha^*_{\vec{P}}(t)\hat{b}^\dagger_{-\vec{P}}\,.
\end{align}
The number of produced pairs for each canonical momentum  \(\vec{P}\) is now given by the transmission probability
\begin{align}
 W_\text{{WKB}}(\vec{P}):=\lim_{t\rightarrow\infty} |\beta_{\vec{P}}(t)|^2
 =\lim_{t\rightarrow\infty}\frac{|R_{\vec{P}}(t)|^2}{1-|R_{\vec{P}}(t)|^2}\approx\lim_{t\rightarrow\infty}|R_{\vec{P}}(t)|^2,
\end{align}
which can be connected to the reflection amplitude \(R_{\vec{P}}= \beta_{\vec{P}}(t)/\alpha_{\vec{P}}(t)\). The time evolution of the reflection amplitude \(R_{\vec{P}}\) becomes the Riccati equation (see, e.g., \cite{Dumlu2011}) 
\begin{align}
 \dot{R}_{\vec{P}}(t)=\frac{\dot{\scE}(t)}{2\scE(t)}\left[\exp\left(-2{\frac{\i}{\hbar}\int^{t}\scE(t')dt'}\right)-(R_{\vec{P}}(t))^2\exp\left(2{\frac{\i}{\hbar}\int^{t}\scE(t')dt'}\right)\right]. \label{eq:Riccati}
\end{align}
For small \(R_{\vec{P}}(t)\)  one can ignore the second non-linear term and approximately solve the Riccati
 equation (\ref{eq:Riccati}) by  integrating
\begin{align}
 \lim_{t\rightarrow \infty} {R}_{\vec{P}}(t)=\int_{-\infty}^\infty \frac{\dot{\scE}(t)}{2\scE(t)}\exp\left(-2{\frac{\i}{\hbar}\int_{-\infty}^{t}\scE(t')dt'}\right)dt . 
\end{align}
This integral is dominated by the neighborhoods of the turning points \(t_p^{\pm}\) defined by
\begin{align}
  \scE(t_p^{\pm})=0 \label{eq:turningpoints}.
\end{align}
It is obvious from \Eqs(\ref{eq:scE}) and (\ref{eq:turningpoints}) that these turning points are momentum dependent and do not take real values. As was discussed in \cite{Popov1968} tunneling paths for time-dependent potentials can be described with the help of imaginary times. This ``imaginary time method'' was applied to the case of pair production in \cite{Popov1971,Popov1972,Popov1973,Marinov1977}.
From the definition of the turning points in \Eq(\ref{eq:turningpoints})
we however find that \(t_p^{\pm}\) are not necessarily purely imaginary, but are found in conjugated pairs in the complex plane. 
The turning points are purely imaginary only for potentials which are odd functions of the time. 
As was already discussed in \cite{Dumlu2011WLI} this is true for the cases which are normally treated in the ``imaginary time method'' namely \(V_1(t)=E_0 t,\,V_1(t)=E_0/\omega \sin(\omega t)\) and \(V_1(t)=E_0/\omega \tanh(\omega t)\). In general it is however necessary to allow complex values for the turning points. \\
The reflection coefficient can be evaluated as the sum over turning points. However the approximation of ignoring the second non-linear term in the Riccati
 equation (\ref{eq:Riccati}) can lead to a wrong prefactor. By also considering this term one finds   (see \cite{Dumlu2011} for details)
\begin{align}
 \lim_{t\rightarrow \infty} {R}_{\vec{P}}(t)\approx\sum_{t_p^{\pm}}\exp\left(-2{\frac{\i}{\hbar}\int_{-\infty}^{t_p^+}\scE(t')dt'}\right).
\end{align}
We can now split the integral in real parts along the imaginary axis and imaginary parts along the real axis. In order to do so we define the real part of the turning points \(s_p\) and the phase integral \(\theta(s,s')\) as 
\begin{align}
 s_p=\text{Re}(t_p^\pm) ,&&&
 \theta(s,s')=\frac{1}{\hbar}\int_{s}^{s'}\scE(t')dt'\label{eq:sp} .
\end{align}
This allows us to introduce a global phase connected to the first turning point \(t_1^\pm\) 
 \begin{align}
\lim_{t\rightarrow \infty} {R}_{\vec{P}}(t)\approx \mathcal{C}_+ \e^{-2\i\theta(-\infty,s_1)}\sum_{t_p^\pm} \e^ {-2\i \theta(s_1,s_p)}\exp\left(-{\frac{2}{\hbar}\int_{s_p}^{t_p^+}\kappa(t')dt'}\right),
\end{align}
where we introduce
\begin{align}
\kappa(t)=\sqrt{-\scE(t)^2}. \label{eq:kappa}
\end{align}
Now the momentum spectrum of the pair creation rate takes the form \cite{Dumlu2011}
\begin{align}
 W_{\text{WKB}}\left(\vec{P}\right)&\approx\lim_{t\rightarrow \infty} |{R}_{\vec{P}}(t)|^2%\\&\approx
 =\sum_{t_p^{\pm}}\e^{-2K(t_p^\pm)}+
 \sum_{t_p^{\pm}\ne t_{p'}^{\pm}}2 
 \cos(2 \theta(s_p,s_{p'}))
 \e^{-K(t_p^\pm)}\e^{-K(t_{p'}^\pm)}, 
\end{align} 
where we introduce the integral
\begin{align}
K(t_p^\pm)=\frac{1}{\hbar}\int_{t_p^-}^{t_p^+} \kappa(t') dt'. \label{eq:thetaK}
\end{align}
As described in \cite{Dumlu2011} the first term is related to the pair production for every distinct pair of turning points whereas the second term is related to the interference between the respective turning points \(t_{p}^\pm\) and \(t_{p'}^\pm\). 
In the following we will concentrate how to best calculate the integral \(K(t_p^-)\) for  the special case that there is one dominant pair of turning points. 
%%%%%%%%%%%%%%%%%%%%%%%%%%%%%%%%%%%%%%%%%%%%%%%%%%%%%%%%%%%%%%%%%%%%%%%%%%%%%%%%%%%%%%%%%%%%%%%%%%%%%%%%%%%%%%%%%%%%%%%%%%%%%
\subsection{Calculation of the integral \texorpdfstring{
\(K(t_p)\)}{K}
}
%%%%%%%%%%%%%%%%%%%%%%%%%%%%%%%%%%%%%%%%%%%%%%%%%%%%%%%%%%%%%%%%%%%%%%%%%%%%%%%%%%%%%%%%%%%%%%%%%%%%%%%%%%%%%%%%%%%%%%%%%%%%%
\label{sec:K}
In this Section we want to introduce analytic tools to calculate the integral \(K(t_p)\) which is defined in \Eq(\ref{eq:thetaK}).
To enhance the compatibility to existing literature (especially the ``imaginary time'' and the world-line instanton method of \cite{Popov1971,Popov1972,Popov2001} and \cite{Dunne2005B,Dunne2006} respectively) we choose to work in natural units in which energies are measured in units of \(m c^2\) and introduce the adiabatic parameter
\begin{align}
 \gamma:=\frac{m\omega c}{e E_0},
\end{align}
as well as the frequency \(\omega\). We can now write the potentials as 
\begin{align}
 V_j(t)=:\frac{f_j(\omega t)}{\gamma} \label{eq:fintro}
\end{align}
for \(j=1,\,2,\,3\). The WKB transmission probability for one pair of turning points is given by 
\begin{align}
 W_{\text{WKB}}\left(\vec{P}\right)=\exp\left(-\pi\frac{E_c}{E_0}G(\vec{P},\gamma)\right) \label{eq:transrate},
\end{align}
where we define the integral 
\begin{align}
 G(\vec{P},\gamma)=\frac{2}{\pi}\frac{E_0}{E_c}K(t_p^\pm). \label{eq:GK}
\end{align}
For the calculation of the integral we change the integration variable by analogy with a change of variable in the one-component case (see, e.g., \cite{Kleinert2008}). In order to do so we define the function 
\begin{align}
 F(\vec{P},t)={\sqrt{ \left[\gamma c P_1-f_1(t)\right]^2+\left[\gamma c P_2-f_2(t)\right]^2+\left[\gamma c P_2-f_3(t)\right]^2}} \label{eq:Fp3}
\end{align}
and with its help change the integration variable to
\begin{align}
\tau=\pm \i \frac{F(\vec{P},\omega t)}{\gamma}, \label{eq:tautwo}
\end{align} 
 such that  \(\tau\left(\vec{P};t^\pm\right)=1\).  Observe that unlike in the one-dimensional case the value of \(\tau\) is \(1\) at both turning points, which would result in a vanishing integral for an integration between these two points. To resolve this problem we have to choose the sign of \(\tau\) carefully. For the integration from \(t_p^-\) to \(s_p\) we choose the negative sign whereas for the integration from \(t=s_p\) to \(t_p^+\) we choose the plus sign. \\
 Here \(s_p\) is the real part of the pair of turning points \(t_p\) defined in \Eq(\ref{eq:sp}).
 Because of the symmetry of the problem these two integrals have the same value and we can summarize them in a single one from \(\tau_0=\tau(s_p)\) to \(\tau=1\). We find 
\begin{align}
G(\vec{P},\gamma)=&\i\frac{\omega}{\gamma^2}\frac{2}{\pi}\int_{\omega t_p^-}^{\omega t_p^+}dt\sqrt{(\gamma c P_j-f_j(t))^2+\gamma^2} \label{eq:Goriginal3}\\
=&\frac{4}{\pi}
\int_{\tau_0}^{1}d\tau \frac{\sqrt{1-\tau^2}}{\mathcal{F}(\vec{P},-\i \gamma \tau)}, \label{eq:Gs3}
\end{align}
where
\begin{align}
 \mathcal{F}(\vec{P},z):=\left.\frac{\partial}{\partial t}F(\vec{P},t)\right|_{t=F^{-1}(\vec{P},z)} \label{eq:F2}
 \end{align}
is the derivative of the function \(F(\vec{P},t)\) defined in \Eq(\ref{eq:Fp3}) re-expressed as a function of \(\tau\). This function is only uniquely defined for one distinct pair of turning points \(t_p^\pm\). If there is more than one of these pairs we would find a function \(\mathcal{F}_{t_p^\pm}(\vec{P},z)\) for each pair \(t_p^\pm\).
%%%%%%%%%%%%%%%%%%%%%%%%%%%%%%%%%%%%%%%%%%%%%%%%%%%%%%%%%%%%%%%%%%%%%%%%%%%%%%%%%%%%%%%%%%%%%%%%%%
\subsection{Pair production rate for time-dependent electric fields}
%%%%%%%%%%%%%%%%%%%%%%%%%%%%%%%%%%%%%%%%%%%%%%%%%%%%%%%%%%%%%%%%%%%%%%%%%%%%%%%%%%%%%%%%%%%%%%%%%%
\label{sec:paircreation}

The pair production rate per volume \(V\) can be found by integrating the momentum spectrum defined in \Eq(\ref{eq:transrate}) over all the possible momenta with respect to energy momentum conservation for multiphoton absorption. 
For \(\gamma\ll1\)  the photon energy spectrum becomes virtually continuous \cite{Popov2001,Popov1973B}. This leads to 
\begin{align}
\frac{\Gamma_\text{WKB}}{V}&\approx\Ds { \hbar \omega} \int  \frac{d^3P}{(2\pi\hbar)^3}   W_{\text{WKB}}(\vec{P})%\\&\approx
=\Ds  { \hbar \omega} \int \frac{d^3P}{(2\pi\hbar)^3}  \exp\left(-\pi\frac{E_c}{E_0}G(\vec{P},\gamma)\right) . \label{eq:WKBratepA}
\end{align}
Here \(\Ds\) is a factor connected to the spin of the particles. For electrons with two spin orientations it is equal to \(2\) \cite{Kleinert2008,Popov1973B}.\\
For comparison with the world-line instanton method of Section \ref{sec:WLI} it is useful to expand \Eq(\ref{eq:GK})  around \(\vec{P}=0\). The explicit calculations for this expansion are performed in \ref{app:series}. We find
\begin{align}
G(\vec{P},\gamma)=G(\vec{0},\gamma)+\frac{1}{2}cP_j G_{jk}(\gamma)cP_k+\cdots, \label{eq:3expansion}
\end{align}
where the linear contributions
\begin{align}
\left. \frac{\partial G(\vec{P},\gamma)}{\partial c P_j}\right|_{\vec{P}=0}=0, \label{eq:Gj}
\end{align}
vanish for \(j=1,2,3\) following from \Eq(\ref{eq:appGj}).
We also define
\begin{align}
G_{jk}(\gamma):=&\left. \frac{\partial^2 G(\vec{P},\gamma)}{\partial (c P_j)\partial (c P_k)}\right|_{\vec{P}=0}\\
\begin{split}
=&\delta_{jk}\frac{4}{\pi}\int_{\tau_0}^1 \frac{1}{\sqrt{1-\tau^2} }\frac{1}{\mathcal{F}(\vec{0},-\i \gamma \tau)}d\tau%\\&
+\frac{1}{\gamma^2}\frac{4}{\pi}\int_{\tau_0}^{1}\frac{1}{\sqrt{1-\tau^2}}\frac{\partial}{\partial \tau}\left(\frac{F_j(-\i \gamma \tau)F_k(-\i \gamma \tau)}{\tau\,\mathcal{F}(-\i \gamma \tau)}\right)d\tau \label{eq:G_jknb}
\end{split}
\end{align}
 for \(j,k=1,2,3\) following from \Eq(\ref{eq:appG_jknb}).   Here we define
\begin{align}
 {F}_j(z):=f_j(F^{-1}(\vec{0},z)). \label{eq:Fj}
\end{align}
After a Gaussian integration the pair creation rate (\ref{eq:WKBratepA})   takes the form
\begin{align}
\frac{\Gamma_\text{WKB}}{V}&\approx\frac{\Gamma_\text{WKB}^{\vec{P}\sim0}}{V}:= \Ds{\hbar \omega} \left(\frac{m c}{2\pi \hbar}\right)^3 \left(\frac{E_0}{E_c}\right)^{3/2} \frac{\exp\left(-\pi\frac{E_c}{E_0}G(\vec{0},\gamma)\right)}{\sqrt{\det [\frac{1}{2}G_{ij}(\gamma)]}}, \label{eq:compare}
\end{align}
which is only true if \(G_{ij}(\gamma)\) is a positive definite matrix.\\
Observe that the expansion discussed here is not always physically justified as can be seen for the 
 example of rotating electrical fields which will be discussed in Sections \ref{sec:rotating} and \ref{sec:modulaterotating}. 
%%%%%%%%%%%%%%%%%%%%%%%%%%%%%%%%%%%%%%%%%%%%%%%%%%%%%%%%%%%%%%%%%%%%%%%%%%%%%%%%%%%%%%%%%%%%%%%%%%
\section{World-line instanton pair creation rate  for electric fields depending on time}
%%%%%%%%%%%%%%%%%%%%%%%%%%%%%%%%%%%%%%%%%%%%%%%%%%%%%%%%%%%%%%%%%%%%%%%%%%%%%%%%%%%%%%%%%%%%%%%%%%
\label{sec:WLI}
Following the ideas presented in \cite{Dunne2005B,Dunne2006} we start from the Euclidean effective action in the world-line path integral formulation \cite{Schubert2001,Kleinert2009}
\begin{align}
 \Gamma_\text{Eucl}[A]=-\int_0^\infty \frac{dT}{T}\e^{-T/\hbar}\int_{x(T)=x(0)} 
 \mathcal{D}x \exp\left[-\frac{1}{\hbar} \int_0^Td\tau\left(m\frac{\dot{x}^2}{4}+\i e A \cdot \dot{x}\right)\right], \label{eq:EuclAct}
\end{align}
 where the path integral \(\int \mathcal{D}x\) is over all closed Euclidean space-time paths \(x^\mu(\tau)\) with period \(T\) in the proper time \(\tau\).
As is well known the pair production rate is connected to the imaginary part of the Minkowski action which can be connected to the Euclidean action (\ref{eq:EuclAct}) for time-dependent fields as \cite{Dunne2006, Kleinert2009}
\begin{align}
 \Gamma=1-\e^{-2 \Im (\Gamma_\text{Mink})}\approx \Im(\Gamma_\text{Mink})=\text{Re}(\Gamma_\text{Eucl}). \label{eq:ratemink} %\label{eq:connection}
 \end{align}
The classical Euler-Lagrange equations take the form
\begin{align}
 m \ddot{x}_\mu=2\i e F_{\mu \nu}(x)\dot{x}_\nu \label{eq:clELE},
\end{align}
where \( F_{\mu\nu}=\partial_\mu A_\nu- \partial_\nu A_\mu\) is the electromagnetic  field strength tensor. For classical solutions  
\begin{align}
 (\dot{x}^\cl)^2=a^2=\text{const.} \label{eq:a^2}
\end{align}
follows directly from the antisymmetry of \(F_{\mu\nu}\) together with  \Eq(\ref{eq:clELE}) by multiplying with \(x_\mu\). Periodic solutions of \Eq (\ref{eq:clELE})  are called world-line instantons. \\
As described in \cite{Dumlu2011WLI} in general these world-line instantons are complex and start and end their trajectories at the semiclassical turning points defined in \Eq(\ref{eq:turningpoints}). If there is more than one distinct pair of turning points the closed trajectories of these instantons may also include interference segments between these pairs. As in the WKB approach we concentrate on potentials with one dominant pair of turning points in this work. For these cases interference effects are negligible.\\
To sum over all closed loops one can choose to fix a point \(x^{(0)}\) on the loop and allow the loop to fluctuate everywhere but at this point. One now expands 
\begin{align}
 x_\mu(\tau)=x_\mu^\cl(\tau)+\eta_\mu(\tau),
\end{align}
where the fluctuations \(\eta_\mu\) vanish at \(x^{(0)}\) 
\begin{align}
 \eta_\mu(0)=\eta_\mu(T)=0.
\end{align}
and follow the Jacobi equations \cite{Morse1960,Kleinert2009}
\begin{align}
 \Lambda_{\mu\nu}\eta_\nu=0, \label{eq:Jacobi}
\end{align}
where the fluctuation operator \(\Lambda_{\mu\nu}\) is defined by
\begin{align}
 \Lambda_{\mu\nu}=-\frac{1}{2}\delta_{\mu\nu}\frac{d^2}{d \tau^2}-\frac{d}{d\tau}Q_{\mu\nu}+Q_{\mu\nu}\frac{d}{d\tau}+R_{\mu\nu}, \label{eq:flucop}
\end{align}
where 
\begin{align}
 Q_{\mu\nu}=c^2\frac{\partial^2 L}{\partial x_\mu \partial \dot{x}_\nu}, &&& 
 R_{\mu\nu}=c^2\frac{\partial^2 L}{\partial x_\mu \partial x_\nu}.
\end{align}
After integrating over \(x^{(0)}\) and  using the Gelfand-Yaglom method following \cite{Kleinert2009} the semiclassical approximation of the path integral \Eq(\ref{eq:EuclAct})  can be written as \cite{Dunne2006,Levit1977,Kleinert2009} 
\begin{align}
\begin{split}
 \Gamma_\text{Eucl}[A]=-\int_0^\infty \frac{dT}{T}&\int \frac{d^4 x^{(0)}}{(\hbar c)^4}\e^{-T/\hbar}\left( \frac{\hbar}{2 \pi T}\right)^2% \\ &\times 
 \e^{\i \theta} \e^{-S[x^\cl](T)/\hbar}\sqrt{\frac{\left|\det\left(\eta^{(\nu)}_{\mu,\,\text{free}}(T)\right)\right|}{\left|\det\left(\eta^{(\nu)}_{\mu}(T)\right)\right|}}, \label{eq:EuclAct2} %\label{eq:semi}
\end{split}
\end{align}
where \(\eta^{(\nu)}_{\mu}(\tau)\) is the solution to the Jacobi equation (\ref{eq:Jacobi}) with the initial conditions
\begin{align}
 \eta_\mu^{(\nu)}(0)=0, &&& \dot{\eta}_\mu^{(\nu)}(0)=\delta_{\mu\nu}. \label{eq:initial}
\end{align}
The free operator is defined by 
\begin{align}
 \Lambda_{\mu\nu}^{\text{free}}=-\frac{1}{2}\delta_{\mu\nu}\frac{d}{d\tau^2}
\end{align}
such that 
\begin{align}
 \det\left(\eta^{(\nu)}_{\mu,\,\text{free}}(T)\right)=T^4
\end{align}
and the phase factor \(e^{\i \theta}\) is determined by the Morse index of the operator \(\Lambda\) \cite{Dunne2006,Levit1977,Morse1960,Kleinert2009}.\\
 This framework was used in \cite{Dunne2006} to calculate the pair creation rate for one component fields depending either on space or on time and generalized to two and three component fields depending on space in \cite{Dunne2006B}. In the following we will study the generalization to three component fields depending on time.\\
With this method one cannot obtain the momentum spectrum of the pair creation rate. This would however be possible if one starts from the world-line path integral
\begin{align}
\begin{split}
 \Gamma_\text{Eucl}[A]=-\int_0^\infty \frac{dT}{T}\e^{-T/\hbar}&\int_{x(T)=x(0)} 
 \mathcal{D}x  \int \mathcal{D}p %\\ &\times 
 \exp\left[-\frac{1}{\hbar} \int_0^Td\tau\left(\dot{x}\cdot p-\frac{1}{2}\left(cp+ceA\right)^2\right)\right], \label{eq:WLImomentum}
\end{split}
 \end{align}
instead of \Eq (\ref{eq:EuclAct}). This has been done for the one component case in \cite{Dumlu2011WLI}. As argued there the version of \cite{Dunne2006} following from the world-line path integral (\ref{eq:EuclAct}) takes the momenta, arising as integration constants, to be zero. The prefactor is produced by the Gaussian integration performed in \Eq(\ref{eq:EuclAct2}). This can be seen as an implicit expansion of the momentum spectrum around \(\vec{P}=0\).  We leave the investigation of the world-line instanton momentum spectrum  in the general three component case for future work.
%%%%%%%%%%%%%%%%%%%%%%%%%%%%%%%%%%%%%%%%%%%%%%%%%%%%%%%%%%%%%%%%%%%%%%%%%%%%%%%%%%%%%%%%%%%%%%%%%%%%%%%%
\subsection{Classical solutions for three-dimensional electrical fields depending on time}
%%%%%%%%%%%%%%%%%%%%%%%%%%%%%%%%%%%%%%%%%%%%%%%%%%%%%%%%%%%%%%%%%%%%%%%%%%%%%%%%%%%%%%%%%%%%%%%%%%%%%5
We start from the four potential (\ref{eq:potential}) in Euclidean form and by analogy with \Eq(\ref{eq:fintro}) use 
\begin{align}
 V_j(x_0)&=-\i\frac{1}{\gamma}\tilde{f}_j\left(\frac{\omega}{c} x_0\right), \label{eq:ftilde}
\end{align}
where \(j=1,2,3\). The classical Euler-Lagrange equations (\ref{eq:clELE}) can be written explicitly as
\begin{align}
  m\ddot{x}_0&=2\frac{eE_0}{c} \tilde{f}_j'\left(\frac{\omega}{c} x_0\right)\dot{x}_j, \label{eq:clEMO2}\\
 m\ddot{x}_j&=-2\frac{eE_0}{c} \tilde{f}_j'\left(\frac{\omega}{c} x_0\right)\dot{x}_0. \label{eq:clEMO1}
\end{align}
The last three equations can be directly integrated
\begin{align}
 \dot{x}^\cl_j=-\frac{2eE_0c^2}{\omega}\tilde{f}_j\left(\frac{\omega}{c} x_0^\cl\right). \label{eq:xj_EM}
\end{align}
Whereas with the help of \Eq(\ref{eq:a^2}) the first one can be solved as
\begin{align}
\dot{x}_0^\cl=\pm  a \sqrt{1-\left(\frac{\tilde{f}_j\left(\frac{\omega}{c} x_0^\cl\right)}{\bar{\gamma}}\right)^2}, \label{eq:dotx0}
\end{align}
where like in \cite{Dunne2006} we define
\begin{align}
\bar{\gamma}=\frac{a\omega} {2 e E_0 c^2}=\frac{a }{2 c}{\gamma}.
\end{align}

\subsection{The fluctuation determinant}
\label{sec:flucdet}
The fluctuation operator (\ref{eq:flucop}) takes the form 
\begin{align}
 \Lambda_{\mu\nu}
&=-\frac{1}{2}\begin{pmatrix}
 \frac{d^2}{d\tau^2}-\frac{d}{d\tau}\left(\frac{\ddot{x}^\cl_j}{\dot{x}_0^\cl}\right)\dot{x}_j^\cl&\frac{\ddot{x}^\cl_m}{\dot{x}_0^\cl}\frac{d}{d\tau} \\
  -\frac{d}{d\tau} \left(\frac{\ddot{x}^\cl_l}{\dot{x}_0^\cl}\right)- \frac{\ddot{x}^\cl_l}{\dot{x}_0^\cl}\frac{d}{d\tau}  & \delta_{lm}\frac{d^2}{d\tau^2}.
\end{pmatrix}
\end{align}
We obtain the 8 independent solutions to the Jacobi equation (\ref{eq:Jacobi})
\begin{align}
\phi^{(0)}(\tau)&=\begin{pmatrix}
                   \dot{x}_0^\cl(\tau) \tilde{I}(\tau)\\
                   \dot{x}_k^\cl(\tau) \tilde{I}(\tau)-\tilde{I}_{k}(\tau)
                  \end{pmatrix},\\
\phi^{(j)}(\tau)&=\begin{pmatrix}
                   \dot{x}_0^\cl(\tau) \tilde{I}_j(\tau)\\
                   \dot{x}_k^\cl(\tau) \tilde{I}_j(\tau)-\tilde{I}_{jk}(\tau)-\tau \delta_{jk}                   
                  \end{pmatrix},\\
\phi^{(3+j)}(\tau)&=\begin{pmatrix}
                   0\\
                   \delta_{jk}
                  \end{pmatrix},\\
\phi^{(7)}(\tau)&=\begin{pmatrix}
                   \dot{x}_0^\cl(\tau)\\
                   \dot{x}_k^\cl(\tau)
                  \end{pmatrix},
\end{align}
where we define the integrals
\begin{align}
 \tilde{I}(\tau)&=\int_0^\tau dt\frac{1}{\left[\dot{x}_0^\cl(t)\right]^2}, \label{eq:tildeint1}&
 \tilde{I}_j(\tau)&=\int_0^\tau dt\frac{\dot{x}_j^\cl(t)}{\left[\dot{x}_0^\cl(t)\right]^2},&
 \tilde{I}_{jk}(\tau)&=\int_0^\tau dt\frac{\dot{x}_j^\cl(t)\dot{x}_k^\cl(t)}{\left[\dot{x}_0^\cl(t)\right]^2}. %\label{eq:tildeint3}
\end{align}
We can now construct the solutions which fulfill the initial conditions (\ref{eq:initial}) as
\begin{align}
\begin{split}
 \eta^{(0)}_\mu(\tau)&=\phi_\mu^{(0)}(\tau)\dot{x}^\cl_0(0),\\
 \eta^{(j)}_\mu(\tau)&=\phi_\mu^{(0)}(\tau)\dot{x}^\cl_j(0)-\phi_\mu^{(j)}(\tau).
\end{split} \label{eq:inisol}
 \end{align}
 Now we want to compute \(\det(\eta_\mu^{(\nu)}(T))\). To simplify the result one however has to be careful about the integrals defined in \Eq(\ref{eq:tildeint1})%-(\ref{eq:tildeint3})
 .
 The reason for this is that the integrals diverge if \(\dot{x}_0(\tau)\) becomes zero in the interval from  \(\tau=0\) to \(\tau=T\). If one however performs the limit \(\lim_{\tau\rightarrow T}\eta_\mu^{(\nu)}(\tau)\) these divergences are canceled. It is possible to separate the divergences into boundary terms with the help of an integration by parts and thus rewrite \(\det(\eta_\mu^{(\nu)}(T))\) in terms of converging integrals
 \begin{align}
  \lim_{\tau\rightarrow T}  \dot{x}^\cl_0(0)\dot{x}^\cl_0(\tau) \tilde{I}(\tau)&=\dot{x}^\cl_0(0)\dot{x}^\cl_0(T)I(T) , \\ 
  \lim_{\tau\rightarrow T}  \left(\dot{x}^\cl_k(\tau) \tilde{I}(\tau)-\tilde{I}_k(\tau)\right)&=\dot{x}^\cl_k(T) I(T)-I_k(T), \\
  \lim_{\tau\rightarrow T}  \left(\tilde{I}_{jk}(\tau)-\dot{x}^\cl_k(\tau) \tilde{I}_j(\tau)\right)&=I_{jk}(T)-\dot{x}^\cl_k(T) I_j(T),
 \end{align}
 where we define the converging integrals
\begin{align}
 I(\tau)&=\int_0^\tau dt\frac{1}{\dot{x}_0^\cl(t)}\frac{\partial}{\partial t}\left(\frac{1}{\ddot{x}_0^\cl(t)}\right),\label{eq:int1} & I_j(\tau)&=\int_0^\tau dt\frac{1}{\dot{x}_0^\cl(t)}\frac{\partial}{\partial t}\left(\frac{\dot{x}_j^\cl(t)}{\ddot{x}_0^\cl(t)}\right),& I_{jk}(\tau)&=\int_0^\tau dt\frac{1}{\dot{x}_0^\cl(t)}\frac{\partial}{\partial t}\left(\frac{\dot{x}_j^\cl(t)\dot{x}_k^\cl(t)}{\ddot{x}_0^\cl(t)}\right). %\label{eq:int3}
\end{align}
Using the periodicity of the classical world-line instantons namely \(\dot{x}^\cl_j(T)=\dot{x}^\cl_j(0)\), for which \(I_j(T)=0\)  follows, we find
\begin{align}
 \eta_\mu^{(\nu)}(T)=\dot{x}^\cl_\mu(0)\dot{x}^\cl_\nu(0)I(T)+I_{ij}(T)+T\delta_{ij},
\end{align}
where \(\mu=(0,i) \und \nu=(0,j)\). So that we can compute the fluctuation determinant
\begin{align}
\det\left(\eta_\mu^{(\nu)}(T)\right)=\left(\dot{x}^\cl_0(0)\right)^2 I(T)\det\left(I_{ij}(T)+T\delta_{ij}\right).
\end{align} 
For the case of the one  component electric field  depending on time (\(\dot{x}_2^\cl(\tau)=\dot{x}_3^\cl(\tau)=0\)) 
studied in \cite{Dunne2006} one finds 
\begin{align}
 \tau+I_{11}(\tau)=\int_0^{\tau}dt+I_{11}(\tau)=\int_{0}^\tau dt\frac{1}{\dot{x}_0^\cl(t)}\frac{\partial}{\partial t} \left(\frac{(\dot{x}_0^\cl(t))^2+(\dot{x}_1^\cl(t))^2}{{\ddot{x}_0^\cl(t)}}\right)=a^2 I(\tau),
\label{eq:Ia2}
 \end{align} 
following from \Eq(\ref{eq:a^2}) and thus we recover
\begin{align}
 \det\left(\eta_\mu^{(\nu)}(T)\right)=\left(\dot{x}^\cl_0(0) I(T)\, T a\right)^2 . \label{eq:fluc1dim}
\end{align}
A factor of \(T^2\),  with respect to  Eq.~(3.22) of \cite{Dunne2006},  stems from the fact that there only the two dimensional (0,1) part of  \(\eta_\mu^{(\nu)}\) is taken into account. This is possible since the (2,3) part is equal to \(T \delta_{ij}\). \\ 
Now we need to calculate the Morse index to determine the phase factor in \Eq(\ref{eq:EuclAct2}). It can be derived either as the number of times the determinant \(\det(\eta_\mu^{(\nu)}(\tau))\) is zero in between \(0\) and \(\tau\) or as the number of negative eigenvalues of the operator \(\Lambda\).  In \cite{Dunne2006} it was stated that for the examples studied there (\(A_1(t)\sim \sin(t) \und A_1(t)\sim \tanh(t)\)) this index is 2, leading to a phase factor of \(-1\). In \ref{app:Morse} we show that this is true for all electric fields with one component depending on time. However we have not been able to prove it for the general three-component case. 
\\ 
We now use 
\begin{align}
 \int d^4  x^{(0)}=\int dx_0(0)dx_1(0)dx_2(0)dx_3(0)= V\int d\tau_0 \dot{x}_0^\cl(0)=V\,\frac{T}{2}\,\dot{x}_0^\cl(0),
\end{align} 
where \(V\) is the 3-space volume. Using (\ref{eq:EuclAct2}) one obtains the semiclassical Euclidean action 
\begin{align}
\Gamma_\text{Eucl}^\text{semi}\approx -\frac{V}{2c^4(2\pi \hbar)^2}\, \e^{\i \theta}\int_0^\infty {dT}\,   \frac{\e^{-[T+S[x^\cl](T)]/\hbar}}{\sqrt{I(T)\det\left(I_{ij}(T)+T\delta_{ij}\right)}} . \label{eq:semiEuclAct}
\end{align}

%%%%%%%%%%%%%%%%%%%%%%%%%%%%%%%%%%%%%%%%%%%%%%%%%%%%%%%%%%%%%%%%%%%%%%%%%%%%%%%%%%%%%%%%%%%%%%%%%%%%%%
\subsection{The exponent}
%%%%%%%%%%%%%%%%%%%%%%%%%%%%%%%%%%%%%%%%%%%%%%%%%%%%%%%%%%%%%%%%%%%%%%%%%%%%%%%%%%%%%%%%%%%%%%%%%%%%%%
\label{sec:exponent}
We now study the exponent in \Eq(\ref{eq:semiEuclAct}) which is proportional to
\begin{align}
 \Delta(T):=S[x^\cl](T)+T.
\end{align}
Using the classical equations of motion (\ref{eq:clEMO1}) and (\ref{eq:clEMO2}) we find
\begin{align}
S[x^\cl](T)=& \int_0^Td\tau\left(m\frac{(\dot{x}^\cl)^2}{4}+\i  \frac{e A}{c} \cdot \dot{x}^\cl \right)%\\
%=& -\frac{1}{4}\frac{a^2}{c^2 }T+\frac{1}{2}\int_0^Td\tau (\dot{x}_0^\cl)^2\\
%=& -\frac{1}{4}\frac{a^2}{c^2 }T+\frac{1}{2}\frac{a^2}{c^2}\int_0^Td\tau  \left(1-\left(\frac{\tilde{f}_j\left(\frac{\omega}{c} x_0^\cl\right)}{\bar{\gamma}}\right)^2\right).
\end{align}
Introducing the function
\begin{align}
 \tilde{F}(t)=\pm\sqrt{\left(\tilde{f}_j(t)\right)^2}, \label{eq:Ftilde}
\end{align}
we change the variable of the integral to 
\begin{align}
y=\tilde{F}\left(\frac{\omega}{c} x_0^\cl\right)/\bar{\gamma}. \label{eq:y}
\end{align}
As result we obtain 
\begin{align}
\Delta(T)=T\left(1-\frac{a(T)^2}{4 c^2}\right)+\frac{ a(T)^2}{4 e E_0 c^3}\pi g(\bar{\gamma}(T))
\end{align}
with
\begin{align}
 g(z)=&\frac{2}{\pi}\int_{-1}^1dy \, \frac{\sqrt{1-y^2}}{\mathcal{F}(z y)} \sgn(y-y_0) %\label{eq:gorig}
 =\frac{4}{\pi}\int_{y_0}^1dy \frac{\sqrt{1-y^2}}{\mathcal{F}(z y)} \label{eq:g},
\end{align}
where
\begin{align}
 \mathcal{F}(z):=\left| \tilde{F}'(\tilde{F}^{-1}(z))\right| \label{eq:calF}
\end{align}
is the derivative of \(F(z)\) re-expressed as a function of \(y\). As discussed in \cite{Dumlu2011WLI}, world-line instantons are closed curves which end and start at the classical turning points which correspond to \(y=\pm1\). This can be motivated since \(\dot{x}_0^\cl(y)=\pm  a \sqrt{1-y^2}\) following from \Eq(\ref{eq:dotx0}) becomes \(0\) at these points such that the interval in between covers half a period. As in the WKB-case when we perform the substitution (\ref{eq:y}) we have to choose the sign in \Eq(\ref{eq:Ftilde})  carefully. We introduce \(y_0=y(s_p)\) by analogy with \(\tau_0\) as discussed in Section \ref{sec:K}.\\
We now want to use a saddle point approximation for the integral over \(T\) in \Eq(\ref{eq:semiEuclAct}) and thus need to calculate 
\begin{align}
\frac{d\Delta(T)}{dT}=
 \left(1-\frac{a(T)^2}{4c^2}\right).
\end{align}
 It follows that the saddle point occurs for \(a(T_c)=2c\) which is equivalent to \(\bar{\gamma}=\gamma\) found in \cite{Dunne2006}.
The second derivative of the exponent at the critical period \(T=T_c\) by analogy with the one component case of \cite{Dunne2006} equals
\begin{align}
\Delta''(T_c)=\left.\frac{d^2\Delta(T)}{dT^2}\right|_{T=T_c}=\frac{1}{2 c^2 I(T_c)},
\end{align}
where   \(I(T)\) is the integral \(I(\tau)\) defined in \Eq(\ref{eq:int1}) over the full period \(T\).

%%%%%%%%%%%%%%%%%%%%%%%%%%%%%%%%%%%%%%%%%%%%%%%%%%%%%%%%%%%%%%%%%%%%%%%%%%%%%%%%%%%%%%%%%%%%%%%%%%%%%%%%%%%%
\subsection{Pair creation rate}
%%%%%%%%%%%%%%%%%%%%%%%%%%%%%%%%%%%%%%%%%%%%%%%%%%%%%%%%%%%%%%%%%%%%%%%%%%%%%%%%%%%%%%%%%%%%%%%%%%%%%%%%%%%%
Now we can use the saddle point approximation for the integral over \(T\) in \Eq(\ref{eq:semiEuclAct})
\begin{align}
\Gamma_\text{Eucl}^\text{semi}\approx& -\e^{\i \theta}\,\frac{V}{2c^4(2\pi \hbar)^2} \,  \sqrt{\frac{\pi \hbar}{2 \Delta''(T_c)}}   \frac{\e^{-\frac{1}{\hbar}\Delta(T_c)}}{\sqrt{I(T_c)\det\left(I_{ij}(T_c)+T_c\delta_{ij}\right)}}
%\\
= -\e^{\i \theta}\,\frac{V}{(2 \sqrt{\pi \hbar}c)^3} \,    \frac{\e^{-\pi \frac{E_c}{ E_0} g(\gamma)}}{\sqrt{\det\left(I_{ij}(T_C)+T_C\delta_{ij}\right)}} .
\end{align}
Using \Eq(\ref{eq:ratemink}) we find that the pair creation rate \(\Gamma\) can be approximated by the imaginary part of the Minkowski action \(\Gamma_\text{Mink}\) and that this in turn is approximately equal to  the semiclassical world-line instanton pair creation rate \(\Gamma_\text{WLI}\)
\begin{align}
 {\Gamma}& \approx \Gamma_\text{WLI}:=-\, \e^{\i \theta} \frac{V}{ (2\pi \hbar c )^3 }\left(\frac{E_0}{E_c}\right)^{3/2}  \, \frac{\e^{-\pi \frac{E_c}{E_0}g(\gamma)}}{  \sqrt{\det\left(P_{ij}(\gamma)+P(\gamma)\delta_{ij}\right)} } , \label{eq:compareWLI}
\end{align}
where we define
\begin{align}
 P(\bar{\gamma}(T)):=\frac{2 }{\pi}\int_{y_0}^{1}d y \frac{1}{ \sqrt{1-y^2}} \frac{1}{\mathcal{F}(\bar{\gamma}(T)y)}=\frac{ c e E_0}{\pi }T \label{eq:T}
\end{align}
by analogy with the one component case of \cite{Dunne2006}. We also define
\begin{align}
P_{jk}(\gamma):=\frac{ceE_0}{\pi}I_{jk}(T_c)=-\frac{1}{\gamma^2}\frac{2}{\pi}\int_{y_0}^1dy\frac{1}{\sqrt{1-y^2}}\frac{\partial}{\partial y}\left(\frac{\tilde{F}_j(\gamma y)\tilde{F}_k(\gamma y)}{y\mathcal{F}(\gamma y)}\right) \label{eq:P_jk}
\end{align}
by using  definition of the integrals, the substitution (\ref{eq:y}) and
\begin{align}
 \ddot{x}_0^\cl(t)=-2eEca\, y  \mathcal{F}(\bar{\gamma}(T)y) \label{eq:ddotx0},
\end{align}
which can be derived from \Eqs(\ref{eq:dotx0}), (\ref{eq:Ftilde}) and (\ref{eq:calF}).
In (\ref{eq:P_jk}) we use
\begin{align}
 \tilde{F}_j(z):=\tilde{f}_j\left(\tilde{F}^{-1}(z)\right). \label{eq:Ftildej}
\end{align}
%%%%%%%%%%%%%%%%%%%%%%%%%%%%%%%%%%%%%%%%%%%%%%%%%%%%%%%%%%%%%%%%%%%%%%%%%%%%%%%%%%%%%%%%%%%%%%%%%%%%%%%%%%%%%%%
\section{Comparison between the WKB and world-line instanton results}
%%%%%%%%%%%%%%%%%%%%%%%%%%%%%%%%%%%%%%%%%%%%%%%%%%%%%%%%%%%%%%%%%%%%%%%%%%%%%%%%%%%%%%%%%%%%%%%%%%%%%%%%%%%%%%%
\label{sec:compare}
We can now compare the results \Eq(\ref{eq:compare}) of the WKB method discussed in Section \ref{sec:WKB}  and \Eq(\ref{eq:compareWLI}) of the world-line instanton approach of Section \ref{sec:WLI}. Observe that \Eq(\ref{eq:compare}) shows the leading order contribution of the pair production rate (\ref{eq:WKBratepA}) where the momentum spectrum was expanded around \(\vec{P}=0\) and that \Eq(\ref{eq:compareWLI}) is the counterpart in the world-line instanton approach. We will show that these two results agree with each other.\\
From the definitions of \(f_j\)  and \(\tilde{f}_j\) in \Eqs(\ref{eq:fintro}) and (\ref{eq:ftilde}), and the definitions of \(F\) and \(\tilde{F}\) in \Eqs(\ref{eq:Fp3}) and (\ref{eq:Ftilde}) respectively we can find
\begin{align}
 F(\vec{0},t)=-\i \tilde{F}(-\i t).
\end{align}
We thus find
\begin{align}
 \mathcal{F}(\vec{0},-\i z)=\mathcal{F}(z),&&&
 F_j(-\i z)=-\i \tilde{F}(z), \label{eq:con}
\end{align}
 which follows from the respective definitions in \Eqs (\ref{eq:F2}), (\ref{eq:Fj}), (\ref{eq:calF})  and (\ref{eq:Ftildej}).\\
Inserting \Eq(\ref{eq:con}) in \Eqs(\ref{eq:Gs3}) and (\ref{eq:G_jknb}) and comparing to \Eqs(\ref{eq:g}), (\ref{eq:T}) and (\ref{eq:P_jk})  we find
\begin{align}
 g(\gamma)&=G(\vec{0},\gamma),\\
 P_{jk}(\gamma)+\delta_{jk} P(\gamma)&=\frac{1}{2} G_{jk}(\gamma).
\end{align}
Thus from \Eqs(\ref{eq:compare}) and (\ref{eq:compareWLI}) we find
\begin{align}
 \Gamma_\text{WKB}^{\vec{P}\sim0}=\Gamma_\text{WLI} \frac{D_s \hbar \omega}{-e^{\i \theta}}.
\end{align}
This means that the world-line instanton result agrees with the expansion of the WKB rate around \(\vec{P}=0\) except for  a factor of \(D_s\hbar \omega\) provided that \(\e^{\i \theta}=-1\). The factor of \(D_s\hbar \omega\) stems from the sum over the virtually continuous energy spectrum performed for the WKB-result in Section \ref{sec:paircreation} and can be added to the world-line instanton result with the same argumentation (see also \cite{Popov2001,Popov1973B}).
In \ref{app:Morse} we show that \(\e^{\i \theta}=-1\) holds for the one-component case but we have not been able to show it generally. This implies that the world-line instanton result agrees with the WKB result where the momentum spectrum was expanded around \(\vec{P}=0\).\\
In the cases where the momentum spectrum is not peaked around \(\vec{P}=0\) the expansion (\ref{eq:3expansion}) around \(\vec{P}=0\) is not a good approximation of \(G(\vec{P},\gamma)\). Accordingly 
the leading order of the WKB result is not given by \Eq(\ref{eq:compare}) but can be derived from the general form in \Eq(\ref{eq:WKBratepA}). Furthermore \Eq(\ref{eq:compareWLI}) derived from the world-line path integral in \Eq(\ref{eq:EuclAct}) does not apply. As discussed in \cite{Dumlu2011WLI} this is due to the fact that the momentum, arising as an integration constant in this framework, was taken to vanish with a Gaussian momentum integration producing the prefactors. As argued before this can be seen as an implicit expansion of the momentum spectrum around \(\vec{P}=0\). It would however be possible to get information about the momentum dependence also in the world-line instanton approach by  making the more general ansatz of \Eq(\ref{eq:WLImomentum}).\\
Note that it is possible to shift the canonical momentum spectrum by adding constant contributions  \(\vec{A}^0\) to the vector potential \(\vec{A}(t)\). The new potential 
\begin{align}
 \vec{A}'(t)=\vec{A}(t)+\vec{A}^0
\end{align}
gives rise to the same electric field. We define the new momentum modes
\begin{align}
 \vec{P}'=\vec{P}+e \vec{A}^0 \label{eq:refchange}.
\end{align}
The momentum spectrum of \(P'_\mu\) is exactly the same as if we had used the original potential \(A_\mu\). Thus the entire momentum spectrum has been shifted by \(-e A^0_\mu\).\\
Note that adding constant contributions to the potential does not change the electric field and thus does not change the pair creation rate. This is clear from \Eq(\ref{eq:WKBratepA}) since the rate arises from an integration over the whole momentum spectrum and thus is unaffected by a shift. Also the physical spectrum of the kinetic momentum
\begin{align}
 \vec{p}=\vec{P}-\vec{A}(t)=\vec{P}'-\vec{A}'(t) \label{eq:kinmom}
\end{align}
remains unaffected.\\
It would be tempting to try to use this property to shift the peak in the momentum spectrum to \(\vec{P}=0\) in order to use the approximation \Eq(\ref{eq:compare}) to avoid the computation of the momentum spectrum (\ref{eq:Gs3}). This would however require \textit{a priori} knowledge of the position of the peak and thus the momentum spectrum i.e.~one would have to compute (\ref{eq:Gs3}).  Additionally the spectrum is not necessarily peaked around a point in the momentum space as is shown by the example of the constant rotating field in Section \ref{sec:rotating} where it is peaked around a circle defined by \((\gamma c P_1)^2+\gamma c P_2)^2=1\). Such that shifting the momentum spectrum does not simplify the problem.
%%%%%%%%%%%%%%%%%%%%%%%%%%%%%%%%%%%%%%%%%%%%%%%%%%%%%%%%%%%%%%%%%%%%%%%%%%%%%%%%%%%%%%%%%%%%%%%%%%%%%%%%%%%%%%%%%%%%%%%%%%%%
\section{Applications}
%%%%%%%%%%%%%%%%%%%%%%%%%%%%%%%%%%%%%%%%%%%%%%%%%%%%%%%%%%%%%%%%%%%%%%%%%%%%%%%%%%%%%%%%%%%%%%%%%%%%%%%%%%%%%%%%%%%%%%%%%%%%
\label{sec:app}
In this Section we use the techniques developed in the previous sections to calculate the pair production rate for   rotating field configurations.
In Section \ref{sec:rotating} we study the two-component case of a constant rotating electric field. Using the WKB-techniques of Section \ref{sec:WKB} we are able to find an analytic expression for the momentum spectrum.
In Section \ref{sec:modulaterotating} we study the two-component problem of a non-constant rotating field. There because of the higher complexity we in general cannot obtain the momentum spectrum analytically.
\\
The analytic solutions in the following sections can be formulated with help of the elliptical integrals \(F(k,\phi)\) and \(E(k,\phi)\) as well as their complete forms \(\K(k)\) and \(\E(k)\) given by
(see \cite{Grad2000} Eq.~8.111.2-3)
\begin{align}
 \text{F}(k,\varphi):=\int_0^\varphi\frac{d \theta}{\sqrt{1-k^2\sin^2(\theta)}},&&\K(k):=F\left(k,\frac{\pi}{2}\right)\label{eq:K},\\
 \text{E}(k,\varphi):=\int_0^\varphi{\sqrt{1-k^2\sin^2(\theta)}}{d \theta},&&
 \E(k):=E\left(k,\frac{\pi}{2}\right)\label{eq:E}.
\end{align}

%%%%%%%%%%%%%%%%%%%%%%%%%%%%%%%%%%%%%%%%%%%%%%%%%%%%%%%%%%%%%%%%%%%%%%%%%%%%%%%%%%%%%%%%%%%%%%%%%%%%%%%%%%%%%%%%%%%%%%%%%%%%%
\subsection{Constant rotating electric field}
%%%%%%%%%%%%%%%%%%%%%%%%%%%%%%%%%%%%%%%%%%%%%%%%%%%%%%%%%%%%%%%%%%%%%%%%%%%%%%%%%%%%%%%%%%%%%%%%%%%%%%%%%%%%%%%%%%%%%%%%%%%%
\label{sec:rotating}
As an example for a two-component electric field we compute the pair creation rate of a rotating electric field. This is not a purely academic example although it is one of the simplest cases. As shown in \cite{Bulanov2003} a circularly-polarized laser wave in a plasma takes exactly this form.
The rotating electric field is described by
\begin{align}
&f_1(t)= \sin( t), && f_2(t)=g\cos( t) \label{eq:circ} ,
 \end{align}
where \(g\) defines the sense of the rotation.\\
Instead of the substitution for the general case discussed in Section \ref{sec:K} it is more convenient to define
\begin{align}
 F(\vec{P},t)=\frac{\sqrt{ \left[\gamma c P_1-f_1(t)\right]^2+\left[\gamma c P_2-f_2(t)\right]^2}}{\sqrt{(cP_3)^2+1}} \label{eq:Fp2}
\end{align}
instead of (\ref{eq:Fp3}). Changing the integration variable according to \Eq(\ref{eq:tautwo}) we find
\begin{align}
G(\vec{P},\gamma)=&\sqrt{1+\left({cP_3}\right)^2}\frac{4}{\pi}
\int_{\tau_0}^{1}d\tau \frac{\sqrt{1-\tau^2}}{\mathcal{F}(\vec{P},-\i \gamma \tau)}, \label{eq:Gs2}
\end{align}
where \(\mathcal{F}(\vec{P},z)\) is defined in \Eq(\ref{eq:F2}).\\
To find \(\mathcal{F}(\vec{P},t)\) we have to solve the equation \(z=F(\vec{P},t)\). Using the definition of \(F(\vec{P},t)\) in \Eq(\ref{eq:Fp2}) we first find the following relationship between sine and cosine 
\begin{align}
 -z^2 (1+(c P_3)^2)+(\gamma c P_\parallel)^2+1=2 \gamma c P_1 \sin(t)+2\gamma g\, c P_2 \cos(t),
\end{align}
where
\begin{align}
P_\parallel:=\sqrt{P_1^2+P_2^2}.
\end{align}
Then by using \(\sin^2(x)+\cos^2(x)=1\) we find the following quadratic equation for \(\cos(t)\)
\begin{align}
  \cos^2(t)- 2 h(z) g\, \frac{ P_2}{P_\parallel} \cos(t)+h(z)^2-\frac{ P_1^2}{P_\parallel^2}=0
\end{align}
with
\begin{align}
 h(z)=\frac{(\gamma c P_\parallel)^2+1-z^2 (1+(c P_3)^2)}{2 \gamma c P_\parallel} ,
\end{align}
which can be solved as
\begin{align}
 \cos(t)=h(z) g \frac{P_2}{P_\parallel}\pm\sqrt{1-h(z)^2}\frac{| P_1|}{P_\parallel}. \label{eq:cost}
\end{align}
We are interested in
\begin{align}
\frac{\partial F(\vec{P},t)}{\partial t}&=-\frac{\left[\gamma c P_1-f_1(t)\right]f_1'(t)+\left[\gamma c P_1-f_2(t)\right]f_2'(t)}{\sqrt{ \left[\gamma c P_1-f_1(t)\right]^2+\left[\gamma c P_2-f_2(t)\right]^2}\sqrt{(cP_3)^2+1}} %\\&
%&=\frac{1}{z}\frac{g\, \gamma cP_2 \sin(t)- \gamma c P_1 \cos(t)}{1+(c P_3)^2}\\
=\mp  \frac{|P_1|}{P_1} \frac{1}{z} \frac{c P_\parallel}{1+(c P_3)^2} \sqrt{1-h(z)^2}.
\end{align}
Using \Eq(\ref{eq:F2}) we find
\begin{align}
 \mathcal{F}(\vec{P},z)=\frac{1}{2}\left|\frac{\sqrt{-\left(z^2-\gamma^2 C_- \right)\left(z^2-\gamma^2 C_+  \right)}}{z}\right| 
\end{align}
with
\begin{align}
  C_{\pm}=\frac{1}{\gamma^2}\frac{(\gamma c P_\parallel\pm1)^2}{1+(c P_3)^2}. \label{eq:Frot}
\end{align}
We find a unique solution (\ref{eq:Frot}) for \Eq(\ref{eq:F2}) such that there are no interference effects.\\
The integral (\ref{eq:Gs2}) takes the form
\begin{align}
G(\vec{P},\gamma)&=\frac{8}{\pi}  \sqrt{1+( cP_3)^2}
\int_{\tau_0}^{1}d\tau  \frac{\sqrt{1-\tau^2} \,\tau}{\sqrt{\left[\tau^2+C_+\right]\left[\tau^2+C_-\right]}}= \frac{4}{\pi} \sqrt{1+( cP_3)^2}\int_{\tau_0^2}^1 d x \frac{\sqrt{1-x}}{\sqrt{(x+C_+)(x+C_-)}} ,
\end{align}
where we change the integration variable to \(x=\tau^2\). This integral can be solved and takes real values for \(\tau_0^2\ge-C_-\) (see \cite{Grad2000} Eq.~3.141.5)\footnote{The other assumption for the integral, namely \(C_-<C_+\), is satisfied for \(P_\parallel>0\).}
\begin{align}
\begin{split}
G(\vec{P},\gamma)=& \frac{8}{\pi}\frac{1}{\gamma}\sqrt{1+( cP_3)^2}\sqrt{1+C_+}%\\&\times
\left[\text{F}\left(\arcsin\left(\sqrt{\frac{1-\tau_0^2}{1+C_-}}\right),\sqrt{\frac{1+C_-}{1+C_+}}\right)\right.\\&\hspace{5cm}
\left.-\text{E}\left(\arcsin\left(\sqrt{\frac{1-\tau_0^2}{1+C_-}}\right),\sqrt{\frac{1+C_-}{1+C_+}}\right)\right], \label{eq:Gtau0}
\end{split}
\end{align}
where we use the elliptic integrals (\ref{eq:K}) and (\ref{eq:E}).\\
Now we have to compute \(\tau_0\). Therefore we need to find the real part of the turning points defined by \Eq(\ref{eq:turningpoints}) or equivalently by \(\tau=\pm1\). We find from \Eq (\ref{eq:cost}) 
\begin{align}
 t_p^\pm=\arccos\left(h(\pm\i\gamma) g \frac{P_2}{P_\parallel}\pm\sqrt{1-h(\pm \i \gamma )^2}\frac{| P_1|}{P_\parallel}\right),
\end{align}
which has the real part 
\begin{align}
 s_p=\text{Re}(t_p^\pm)=\arcsin\left(\frac{P_1}{P_\parallel}\right)=\arccos\left(g\frac{P_2}{P_\parallel}\right).
\end{align}
This leads to 
\begin{align}
\begin{split}
 \tau_0^2&=\left[\tau\left(s_p\right)\right]^2=-\frac{1}{\gamma^2}\frac{{ \left[\gamma c P_1-f_1(s_p)\right]^2+\left[\gamma c P_2-f_2(s_p)\right]^2}}{{(c P_3)^2+1}}%\\&
 =-\frac{(\gamma c P_\parallel)^2-2 \gamma  c P_\parallel  +1}{(c P_3)^2+1}=-C_-,
\end{split}
 \end{align}
which is used to simplify \Eq (\ref{eq:Gtau0}) to
\begin{align}\,G(\vec{P},\gamma)&=\frac{8}{\pi}  \frac{1}{\gamma} c \sqrt{P_+^2+P_3^2}\left[\K\left(\sqrt{\frac{P_-^2+P_3^2}{P_+^2+P_3^2}}\right)-\E\left(\sqrt{\frac{P_-^2+P_3^2}{P_+^2+P_3^2}}\right)\right] \label{eq:Gfinalunex},
\end{align}
where
\begin{align}
\gamma c P_\pm:=\sqrt{\left(\gamma c P_\parallel \pm 1\right)^2+\gamma^2}. \label{eq:ppm}
\end{align}
We compare the result \Eq(\ref{eq:Gfinalunex}) to Eq.~(28) of \cite{Bulanov2003} . Using the relations between our variables and theirs which are \( p_x=P_3,\, p_z=P_1\) we find
\begin{align}
\left. G(\vec{P},\gamma)\right|_{P_1=p_z,\,P_2=0,\,P_3=p_x}=g(\gamma)+c_x(\gamma) \frac{p_x^2}{m^2}+\O(p_z^2,p_x^4),
\end{align}
this means the result is the same for \(P_2=0\). This is due to the fact that they use the purely ``imaginary time'' picture in contrast to the ``complex time'' we use. Because of  \(\sin(\i x)=i\sinh(x)\) and \(\cos(\i x)=\cosh(x)\) they have to set the momentum in the direction in which the potential is given by a sine equal to zero to get a real result. As discussed in Section \ref{sec:transmission} and in \cite{Dumlu2011WLI} using a purely imaginary time is only feasible for potentials which are odd functions of the time \(t\). \\
\begin{figure}

\includegraphics[scale=0.5,keepaspectratio=true]{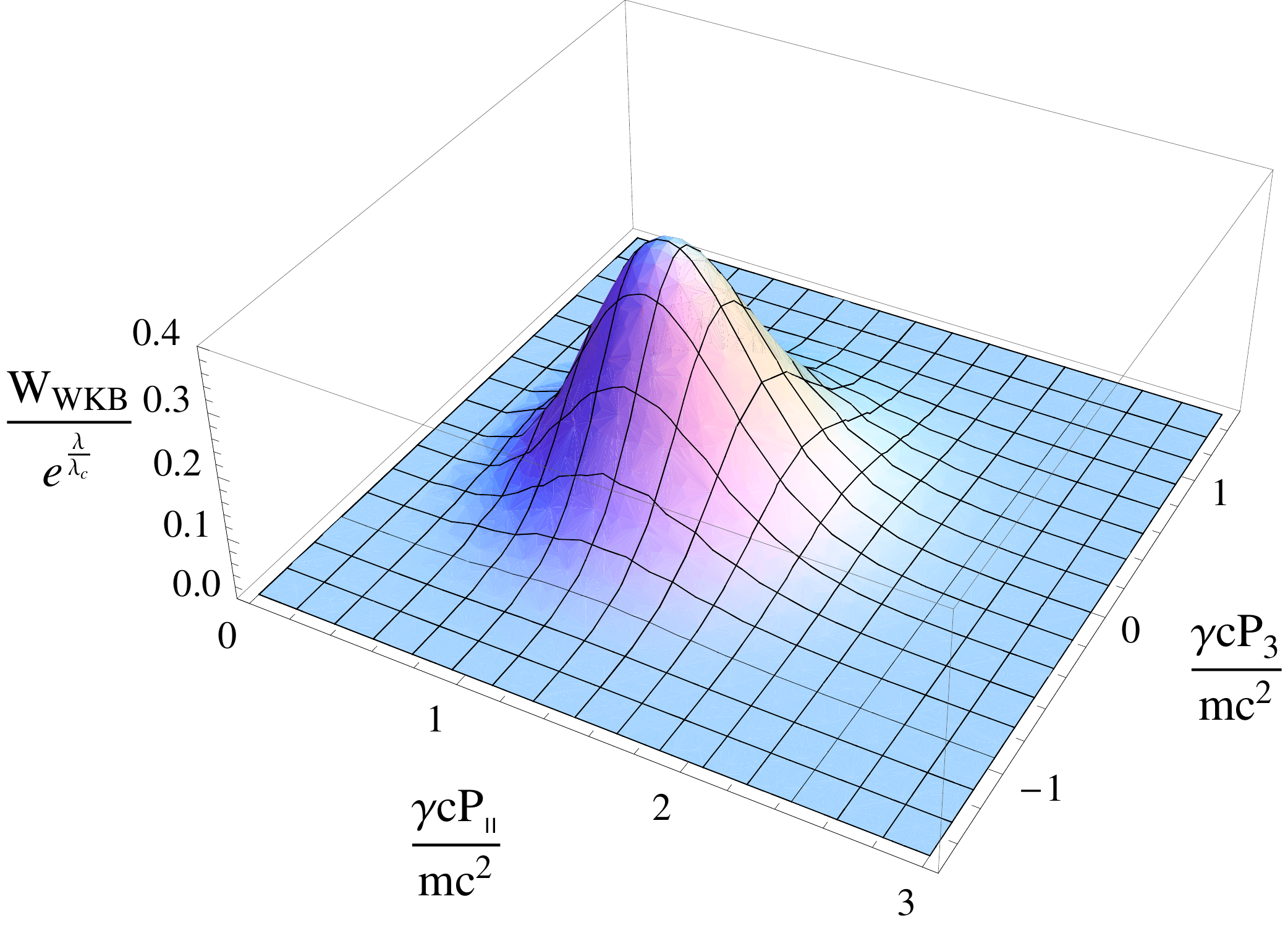}
 \caption{Momentum spectrum \(W_\text{WKB}\) (\ref{eq:transrate}) of the constant rotating field (\ref{eq:circ}) for  \(\gamma=0.5\) (right) plotted as a function of the momenta \(P_3\) and \(P_\parallel\). It is shown that the spectrum is peaked around \(P_3=0\) while for the parallel momentum it is peaked around \(P_\parallel=mc/\gamma\).  } 
 \label{fig:circmomspec}
\end{figure}
If we try to expand in \(P_1,P_2\) around \(0\) as described in Section \ref{sec:paircreation} we run into problems. For \(P_1=P_2=0\) we find \(P_+=P_-\) and thus \(G(\vec{0},\gamma)\) diverges and the pair production rate becomes zero for \(\vec{P}=0\). This means in a constant rotating field no pairs are produced with momentum \(\vec{P}=0\).\\
As discussed in Section \ref{sec:compare} the world-line instanton approach does not apply here since the spectrum is not peaked around \(\vec{P}=0\). Observe that if we want to construct the world-line instanton for \(\dot{x}_
0\) by solving the equation of motion (\ref{eq:dotx0}) we find 
\begin{align}
x_0^\cl(\tau)=\pm  a \sqrt{1-\frac{1}{\bar{\gamma}^2}}\tau+C,
\end{align}
which is not periodic and thus we are not able to construct a world-line instanton for this particular problem.\\
We can however perform the expansion
\begin{align}
G(\vec{P},\gamma)&=\left.G(\vec{P},\gamma)\right|_{P_3=0}+G_3(P_1,P_2,\gamma) (cP_3)^2+\cdots, \label{eq:2expansion}
\end{align}
in \(P_3\) to find 
\begin{align}
 \left. G(\vec{P},\gamma)\right|_{P_3=0}
 &=\frac{8}{\pi}\frac{1}{\gamma} c P_+\left[\K\left(\frac{P_-}{P_+}\right)-\E\left(\frac{P_-}{P_+}\right)\right],\label{eq:G0circ}\\
G_3(P_1,P_2,\gamma)&=\frac{4}{\pi}\frac{1}{\gamma}  \frac{1}{c P_+}\K\left(\frac{P_-}{P_+}\right).\label{eq:G3circ}
\end{align}
This expansion is senseful since \(E_c\) is proportional to \(1/\hbar\) and therefore  the exponential in \Eq(\ref{eq:transrate}) restricts the perpendicular momentum \(P_3\) to be of the order \(\sqrt{\hbar}\) as was discussed for the one-component space-dependent case in \cite{Kleinert2008}.\\
We find that \(G(\vec{P},\gamma)\) is peaked around \( \gamma c P_\parallel=1\). This is in accordance with the results of \cite{Blinne2013} where the momentum spectrum of a rotating pulse is studied numerically. For a high number of rotation cycles per pulse their momentum spectrum is also peaked around a circle of fixed \( P_\parallel\).\\ 
For \(\gamma\ll 1\) an expansion around \( \gamma c P_\parallel=1\) is equal to an expansion for small \(P_-\) which follows from the definition in \Eq (\ref{eq:ppm}).  
 So that \Eqs (\ref{eq:G0circ}) and (\ref{eq:G3circ}) can be expanded for \(P_-\) around \(0\) leading to 
\begin{align}
 \left. G(\vec{P},\gamma)\right|_{P_3=0}&=1+\frac{1}{\gamma^2}(\gamma cP_\parallel-1)^2+\O\left(P_-^4\right),\\
G_3(P_1,P_2,\gamma)&=1+\O\left(P_-^2\right).
\end{align}
Performing the Gaussian integrals over \(P_\parallel\) and \(P_3\) in \Eq (\ref{eq:WKBratepA}) for this approximation we find
\begin{align}
\begin{split}
 \frac{\Gamma_\text{WKB}}{V}&\overset{\hphantom{\gamma\rightarrow0}}\approx
 \Ds  {\hbar \omega}  {\left(\frac{m c}{2\pi\hbar}\right)^3} \exp\left(-\pi\frac{E_c}{E_0}\right)%\\& \times
\left(\left(\frac{E_0}{E_c}\right)^{3/2}\exp\left(-\pi\frac{E_c}{E_0}\frac{1}{\gamma^2}\right)+\frac\pi\gamma  \frac{E}{E_c}\left[1+\text{Erf}\left(\sqrt{\pi\frac{E_c}{E_0}}\frac{1}{\gamma}\right)\right]\right) %\label{eq:ppratecircular} 
\\
&\overset{\gamma\rightarrow0}\approx {\Ds} \left(\frac{m c}{2\pi\hbar}\right)^3 2{\pi}  
 \left(\frac{E_0}{E_c}\right)^2\exp\left(-\pi\frac{E_c}{E_0}\right). \label{eq:pprateconst}
\end{split} 
 \end{align}
 For \(\gamma\rightarrow0\) the pair production rate is equal to the one of the constant field. As already mentioned in \cite{Bulanov2003} this is due to the fact that this limit is equivalent to the limit \(\omega\rightarrow 0\) in which the electric field becomes the constant one. \\
 We can compare the result obtained by numerically integrating \Eq(\ref{eq:WKBratepA}) for \(G(\vec{P},\gamma)\) given by \Eq(\ref{eq:Gfinalunex}) with the one of the constant field given by \Eq(\ref{eq:pprateconst}) for the same amplitude \(E_0\).  We find that the ratio between the two can increase several orders of magnitude with increasing frequency \(\omega\)  (see Fig. 4).  This is in accordance with the results found for rotating pulses in Ref.~\cite{Blinne2013}. We additionally find that this increase is significantly higher if the amplitude of \(E_0\) of the fields is smaller.
As discussed in \cite{Blinne2013} this can be qualitatively understood by the fact the parameter \(\gamma\) is proportional to \(\omega\). With increasing \(\gamma\) the pair creation rate increases due to the onset of multiphoton pair production. Since \(\gamma\) is inversely proportional to \(E_0\) this effect is weakened for higher amplitude field strengths (see also \cite{Kleinert2013} for a similar discussion). 
 \begin{figure}
 \centering
 \includegraphics[scale=0.5,keepaspectratio=true]{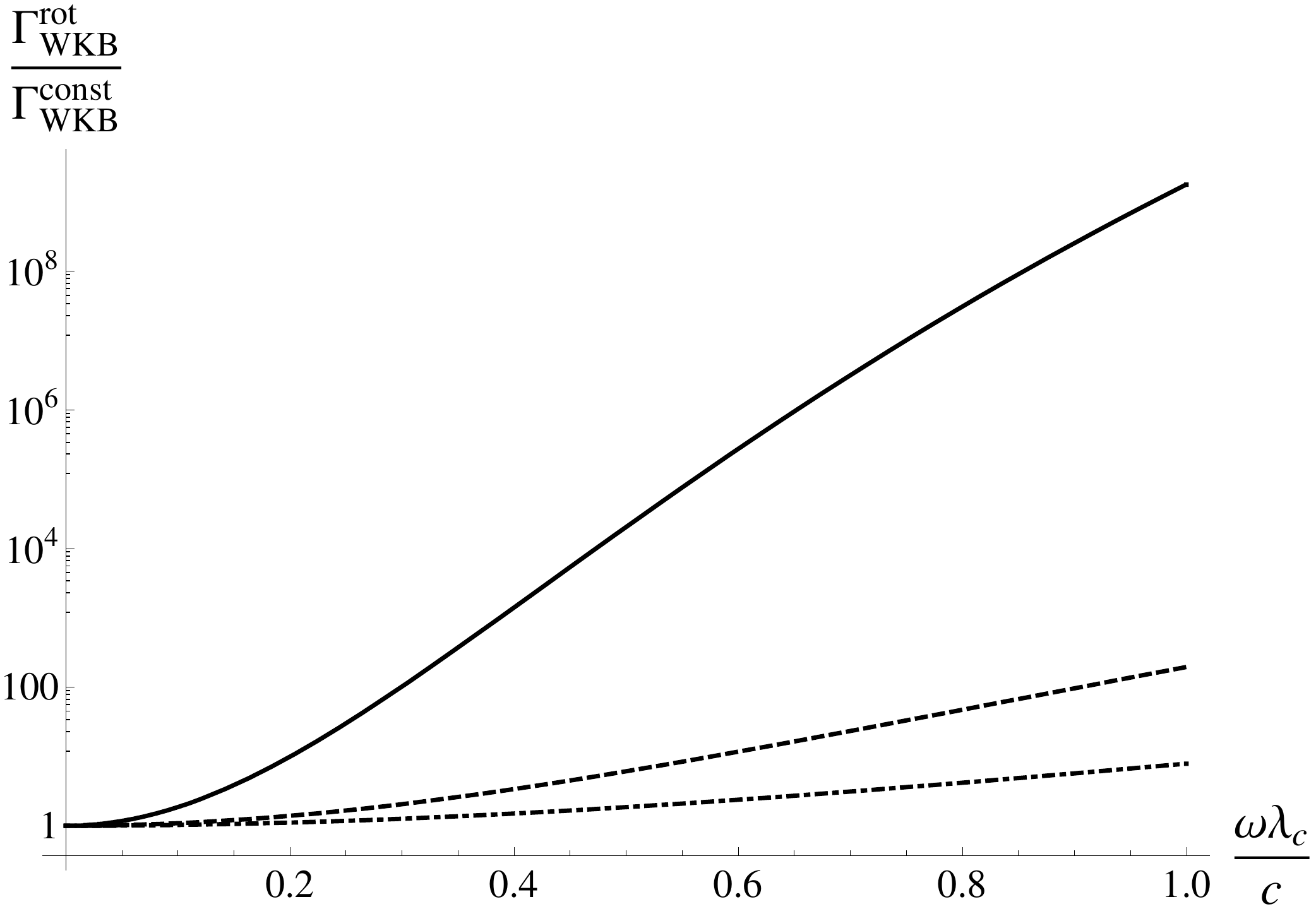}
 % GComparison.pdf: 600x385 pixel, 72dpi, 21.17x13.58 cm, bb=0 0 600 385
 \caption{Ratio of the pair creation rate \Eq(\ref{eq:WKBratepA}) for the constant rotating electric field given by \Eq(\ref{eq:circ}) and the pair creation rate for the constant electric field given by \Eq(\ref{eq:pprateconst}) plotted as a function of the frequency \(\omega\) for \(E_0=0.05 E_c\) (solid) \(E_0=0.1 E_c\) (dashed) and \(E_0=0.15 E_c\) (dot-dashed). The pair creation rate increases with increasing frequency. Additionally this increase is smaller the bigger the amplitude \(E_0\) is.  }
 \label{fig:compare2}
 \end{figure}
 
%%%%%%%%%%%%%%%%%%%%%%%%%%%%%%%%%%%%%%%%%%%%%%%%%%%%%%%%%%%%%%%%%%%%%%%%%%%%%%%%%%%%%%%%%%%%%%%%%%%%%%%%%%%%%%%%%%%%%%%%%%%%%
\subsection{Non-constant rotating field}
%%%%%%%%%%%%%%%%%%%%%%%%%%%%%%%%%%%%%%%%%%%%%%%%%%%%%%%%%%%%%%%%%%%%%%%%%%%%%%%%%%%%%%%%%%%%%%%%%%%%%%%%%%%%%%%%%%%%%%%%%%%%
\label{sec:modulaterotating}
To underline the peculiarities of the expansion (\ref{eq:3expansion})  we study the case of the potential \(A(t)=k(\omega t)/({\gamma e}) \) which rotates with a frequency \(\Omega\), described by
\begin{align}
&f_1(\omega t)=k(\omega t) \sin(\Omega t) , && f_2(\omega t)=k(\omega t) \cos(\Omega t) . 
 \end{align}
It is in general not easy to calculate \(G(\vec{P},\gamma)\) following \Eq (\ref{eq:Gs2}) since it is non-trivial to invert \(F(\vec{P},t)\). However looking at the expansion around \(\vec{P}=0\) described in Section \ref{sec:paircreation} we find
\begin{align}
 F(\vec{0},t)=k(t) \label{eq:Fh}
\end{align}
and  thus \(G(\vec{0},\gamma)\) for the rotating electric field is the same as for the non-rotating with potential \(A_1(t)= k(\omega t)/({\gamma e})\). However as we have seen in Section \ref{sec:rotating} the momentum spectrum is not necessarily peaked around \(\vec{P}=0\) such that the approximation made for the pair creation rate (\ref{eq:compare}) is generally not correct. \\
This point will be further illustrated by the example
\begin{align}
 k(t)=\sin(t).
\end{align}
For this case \(G(\vec{0},\gamma)\) and \(G_{33}(\gamma)\) are given by 
\begin{align}
G(\vec{0},\gamma)&=\frac{4}{\pi}\frac{\sqrt{\gamma^2+1}}{\gamma^2}\left[\K\left(\sqrt{\frac{\gamma^2}{1+\gamma^2}}\right)-\E\left(\sqrt{\frac{\gamma^2}{1+\gamma^2}}\right)\right], \label{eq:G0osci}\\
G_{33}(\gamma)&=\frac{4}{\pi}\frac{1}{\sqrt{1+\gamma^2}}\K\left(\sqrt{\frac{\gamma^2}{1+\gamma^2}}\right). \label{eq:G33osci}
\end{align}
Note that the necessary calculations are analogous to those for an oscillating one-component electric field because of \Eq(\ref{eq:Fh}). Additionally \(G_{13}=G_{23}=0\)  follow from \(f_3(t)=0\).\\
To calculate the rest of the integrals \(G_{jk}(\gamma)\) we need the functions \(F_j\)  (\ref{eq:Fj})  given by
\begin{align}
 F_1(z)=z \sin(\sigma \arcsin(z)), &&& F_2(z)=z \cos(\sigma \arcsin(z)),
\end{align}
where the ratio of the two frequencies \(\sigma:=\Omega/\omega\). If the ratio is an integer \(\sigma=n\) one can calculate \(\mathcal{F}(\vec{P},z)\) and the integrals \(G_{ij}\)  defined in \Eqs(\ref{eq:F2}) and (\ref{eq:G_jknb}) respectively  analytically by using the following identities for sine and cosine functions with multiple angles (see \cite{Grad2000} Eqs.~1.331.1 and 1.331.3)
\begin{align}
 \sin(n x)&=\sum_{j=0}^{\left\lfloor\frac{n-1}{2}\right\rfloor} (-1)^j
 \begin{pmatrix}
 n\\2j+1  
 \end{pmatrix}
\sin^{2j+1}(x)\cos^{n-2j-1}(x),\\
\cos(n x)&=\sum_{j=0}^{\left\lfloor\frac{n}{2}\right\rfloor} (-1)^j
 \begin{pmatrix}
 n\\2j  
 \end{pmatrix}
\sin^{2j}(x)\cos^{n-2j}(x).
\end{align}
If one performs this tedious but simple computations one finds \(G_{12}(\gamma)=0\) and \(G_{11}(\gamma)\) and \(G_{22}(\gamma)\) being combinations of elliptic integrals. However \(G_{11}(\gamma)>0\) and \(G_{22}(\gamma)<0\). This would lead to \(\det(G_{ij}(\gamma))<0\) and thus \Eq(\ref{eq:compare}) would give an imaginary result, which is clearly unphysical . However the Gaussian integral performed to get \Eq(\ref{eq:compare}) is only correct for \(G_{ij}\) being a positive definite matrix. 
So like in Section \ref{sec:rotating} where the terms of the expansion (\ref{eq:3expansion}) diverge we cannot use it to simplify the calculation of the pair creation rate in this case.\\
For \(\sigma=1\) we can show that this is connected to the fact that the momentum spectrum is not centered around \(\vec{P}=0\). In this case we find (see \cite{Grad2000} Eqs. 1.321.1 and 1.333.1)
\begin{align}
 f_1(t)=\sin^2(t)=\frac{1}{2}(1-\cos(2t)),&& f_2=\sin(t)\cos(t)=\frac{1}{2}\sin(2t).
\end{align}
Since adding a constant part to the vector potential has no influence on the electric field, this is analogous to the case of the constant rotating electric field discussed in Section \ref{sec:rotating} with twice the frequency. By shifting the momentum spectrum  \(P_1\rightarrow P_1'\) with \(P_1':=P_1-1/(2\gamma c) \) and defining \(P_\parallel':=\sqrt{(P_1')^2+(P_2)^2}\) the situation is the same as in  Section \ref{sec:rotating} and the spectrum is peaked around \(\gamma cP_\parallel'=1\). As discussed in Section \ref{sec:compare} this shift has no influence on the physical spectrum of the kinetic momentum (\ref{eq:kinmom}).
 
\section{Conclusions and Remarks}
\label{sec:conclusions}
In this article we generalize the analytic methods of the semiclassical WKB-approach and the world-line instanton approach of \cite{Dunne2006} to calculate the pair production rate of time-dependent electric fields to the case of general three-component fields. For the WKB-approach we obtain the momentum spectrum of the produced pairs. We show that if this spectrum is expanded around \(\vec{P}=0\) the results of the two methods are the same.\\
The momentum spectrum is usually peaked around \(\vec{P}=0\) for the examples of one-component fields studied in the literature (see, e.g., \cite{Dunne2006}). Thus the expansion around \(\vec{P}=0\) presents a good approximation. However this situation changes if one goes to the case of two-component fields. By looking at rotating electric fields we find that their momentum spectra are not peaked around  \(\vec{P}=0\).\\
If the momentum spectrum is not peaked around \(\vec{P}=0\) one can not use the expanded WKB result since it does not present a good approximation. Also the world-line instanton method of \cite{Dunne2006} and the generalized form presented here implicitly require the momentum spectrum to be peaked around \(\vec{P}=0\). This implies that it is not appropriate to calculate the pair production rate for cases where the momentum spectrum is not peaked around \(\vec{P}=0\) in the form discussed here.\\
However this can possibly be solved  in the framework of the modified world-line instanton approach of \cite{Dumlu2011WLI}.\\
In this first investigation we ignored the effects of interference which can play an important role, if there is more than one pair of semiclassical turning points. It has been shown, in \cite{Dumlu2011WLI} that the interference effect is the same in the WKB-approach and the world-line instanton method for the case of electric fields with one component. The investigation of this in the general three-component case is left for future work.  
\\
Rotating field configurations such as the one studied here are of interest since they are related to circularly-polarized laser waves. A circularly-polarized wave in medium can be described by a rotating electric field, since it is possible to make a transformation into the co-moving Lorentz frame (see, e.g.,~\cite{Bulanov2003}).\\
Recently it has become obvious that the pair production rate of lasers depends sensitively on the pulse shape  \cite{Schutzhold2008,Dunne2009,Bell2008,DiPiazza2009,Monin2010,Monin2010B,Heinzl2010,Bulanov2010}. For the design of feasible experiments to directly  measure pair production it is therefore of interest to find a pulse profile which enhances this process. Obviously for complicated laser pulse profiles the calculation  has to be done numerically. The development of semiclassical analytical methods discussed in this article certainly  helps to provide some physical intuition for these numerical simulations. 

%%%%%%%%%%%%%%%%%%%%%%%%%%%
\section*{Acknowledgements}
The authors thank Hagen Kleinert and Remo Ruffini for many fruitful discussions. We appreciate discussions on the issue of pair production using lasers with Christoph H. Keitel and Antonino Di Piazza.
ES is supported by the Erasmus Mundus Joint Doctorate Program by Grant Number 2012-1710 from the EACEA of the European Commission.
%%%%%%%%%%%%%%%%%%%%%%%%%%%

\appendix

\section{Computation of the Taylor series terms of \texorpdfstring{
\(G(\vec{P},\gamma)\)
}{G}
}
\label{app:series}
In this Appendix we will summarize the calculation of the terms of the Taylor series presented in Section \ref{sec:paircreation}.
The result (\ref{eq:Gj}) can be achieved from the form (\ref{eq:Goriginal3}) of \(G(\vec{P},\gamma)\) with the help of 
\begin{align}
\left. \frac{\partial G(\vec{P},\gamma)}{\partial c P_j}\right|_{\vec{P}=0}=-\frac{1}{\gamma}\frac{2}{\pi}\left(\int_{\tau_0}^{1}-\int_{-1}^{\tau_0}\right)\frac{F_j(-\i \gamma \tau)}{\sqrt{1-\tau^2}\mathcal{F}(\vec{0},-\i \gamma \tau)}d\tau=0, \label{eq:appGj}
\end{align}
Observe that the boundaries of the integral in \Eq(\ref{eq:Goriginal3}), i.e.  \(t_p^+\) and \(t_p^-\) defined in \Eq(\ref{eq:turningpoints}), are functions of \(\vec{P}\) but that the term related to them is proportional to
\begin{align}
 \kappa(t_p^+)\frac{\partial t_p^+(\vec{P})}{\partial P_1}-\kappa( t_p^-)\frac{\partial t_p^-(\vec{P})}{\partial P_1}=0
\end{align}
and thus vanishes because of \Eqs(\ref{eq:turningpoints}) and (\ref{eq:kappa}). \\
\Eq(\ref{eq:G_jknb}) can be found with the help of
\begin{align}
G_{jk}(\gamma)=&\left. \frac{\partial^2 G(\vec{P},\gamma)}{\partial (c P_j)\partial (c P_k)}\right|_{\vec{P}=0}\\
\begin{split}
=&\delta_{jk}\frac{4}{\pi}\int_{\tau_0}^1 \frac{1}{\sqrt{1-\tau^2} }\frac{1}{\mathcal{F}(\vec{0},-\i \gamma \tau)}d\tau%\\&
-\frac{1}{\gamma^2}\frac{4}{\pi}\int_{\tau_0}^{1}\frac{F_j(-\i \gamma \tau)F_k(-\i \gamma \tau)}{(1-\tau^2)^{3/2}\mathcal{F}(\vec{0},-\i \gamma \tau)}d\tau %\\&
+\frac{1}{\gamma^2}\frac{4}{\pi}\left[\frac{F_j(-\i \gamma \tau)F_k(-\i \gamma \tau)}{\sqrt{1-\tau^2}\,\tau\,\mathcal{F}(-\i \gamma \tau)}\right]_{\tau_0}^1\label{eq:G_jk}
\end{split}
\\
\begin{split}
=&\delta_{jk}\frac{4}{\pi}\int_{\tau_0}^1 \frac{1}{\sqrt{1-\tau^2} }\frac{1}{\mathcal{F}(\vec{0},-\i \gamma \tau)}d\tau%\\&
+\frac{1}{\gamma^2}\frac{4}{\pi}\int_{\tau_0}^{1}\frac{1}{\sqrt{1-\tau^2}}\frac{\partial}{\partial \tau}\left(\frac{F_j(-\i \gamma \tau)F_k(-\i \gamma \tau)}{\tau\,\mathcal{F}(-\i \gamma \tau)}\right)d\tau \label{eq:appG_jknb}
\end{split}
\end{align} 
for \(j,k=1,2,3\). The corresponding boundary term is proportional to 
\begin{align}
 \frac{f_j(\omega t_p^+)}{\kappa(t_p^+)}\frac{\partial t_p^+(\vec{P})}{\partial P_k}-\frac{f_j(\omega t_p^-)}{\kappa(t_p^-)}\frac{\partial t_p^-(\vec{P})}{\partial P_k} \label{eq:boundaryterm}
\end{align}
and does not vanish but reduces to the last term in \Eq(\ref{eq:G_jk}) by using 
\begin{align}
\left. \frac{\partial c P_j}{\partial t}\right|_{t=F^{-1}(\vec{0},z)}=-\frac{1}{\gamma}\frac{\mathcal{F}(\vec{0},z)z}{F_j(\vec{0},z)},
\end{align}
which can be deduced by solving \Eq(\ref{eq:Fp3}) for \(P_j\). The term (\ref{eq:boundaryterm}) is important since it cancels the divergence in the second integral  in \Eq(\ref{eq:G_jk}).

\section{Morse Index}
\label{app:Morse}
The Morse index can be determined either by the number of negative eigenvalues of the fluctuation operator \(\Lambda\) (\ref{eq:flucop}) or the number of times \(\det(\eta_\mu^{(\nu)}(\tau))\) vanishes for \(\tau\) in the interval between \(0 \und T\) where \(\eta_\mu^{(\nu)}(\tau)\) are the solutions to the initial value problem (\ref{eq:initial}) of the Jacobi equation (\ref{eq:Jacobi}) \cite{Dunne2006,Levit1977,Morse1960,Kleinert2009}. We choose the latter method to determine the index because we can readily compute the determinant from the solutions (\ref{eq:inisol}) obtained in Section \ref{sec:flucdet}
\begin{align}
 \det\left(\eta_\mu^{(\nu)}(\tau) \right)=\dot{x}_0^\cl(0)\dot{x}_0^\cl(\tau)\tilde{I}(\tau) \det\left(\tilde{I}_{kl}(\tau)+\tau \delta_{kl}-\frac{\tilde{I}_k(\tau)\tilde{I}_l(\tau)}{\tilde{I}(\tau)}\right). \label{eq:deteta}
\end{align}
Following from the classical solution (\ref{eq:dotx0}) and using the substitution (\ref{eq:y}) we find
\begin{align}
\dot{x}_0^\cl(y)=\pm  a \sqrt{1-y^2}.
\end{align}
Since the interval for \(\tau\) from \(0\) to \(T\) is equivalent to twice the one for \(y\) from \(-1\) to \(1\) we find that \(\dot{x}_0^\cl(\tau)\) becomes zero twice namely for \(\tau(y=\pm1)\). This means that the Morse index is at least two.\\
For the case of the one-component electric fields with  (\(\tilde{I}_{2}(\tau)=\tilde{I}_{3}(\tau)=\tilde{I}_{2j}(\tau)=\tilde{I}_{3j}(\tau)=0\)) we show that the Morse index is exactly two. In this case (\ref{eq:deteta}) takes the form
\begin{align}
  \det\left(\eta_\mu^{(\nu)}(\tau) \right)&=\dot{x}_0^\cl(0)\dot{x}_0^\cl(\tau)\tau^2\left(\tilde{I}(\tau)[\tilde{I}_{11}(\tau)+\tau] -{(\tilde{I}_1(\tau))^2}\right)\\
  &=\dot{x}_0^\cl(0)\dot{x}_0^\cl(\tau)\tau^2\left[a\tilde{I}(\tau) -{\tilde{I}_1(\tau)}\right]\left[a\tilde{I}(\tau) +{\tilde{I}_1(\tau)}\right], \label{eq:deteta1}
\end{align}
where we use \Eq(\ref{eq:Ia2}). Substituting (\ref{eq:y}) into the integrals (\ref{eq:tildeint1})%-(\ref{eq:tildeint3})
we find
\begin{align}
a\tilde{I}(\tau)\pm \tilde{I}_1(\tau)=\frac{1}{2eE_0ca}\int_{y(0)}^{y(\tau)}d y \frac{1\mp y}{ (1-y^2)^{3/2}} \frac{1}{\mathcal{F}(\bar{\gamma}(T) y)}, \label{eq:Ia+I1}
\end{align}
where \Eq (\ref{eq:xj_EM}) is used to find \(\dot{x}_1^\cl(\tau)=-a y\). Since \(-1<y(\tau)<1\), the integrand is always positive. This means that the integral (\ref{eq:Ia+I1}) is only zero for \(\tau=0\). This implies the zero points of (\ref{eq:deteta1}) are located at \(\tau(y=\pm1)\) and \(\tau=0\). Since these points are the same, the determinant becomes zero twice for \(0<\tau<T\), i.e.~the Morse index  \(\theta=2\) for the case of one-component electric fields depending on time.

\bibliographystyle{elsarticle-num}

\end{document}